# Sequential Self-Propelled Morphology Transitions of Nanoscale Condensates Diversify the Jumping-Droplet Condensation

*Shan Gao*[1,3†], *Jian Qu*[1†], *Zhichun Liu*[2,*] and *Weigang Ma*[3,*]

[1]School of Energy and Power Engineering, Jiangsu University, Zhenjiang 212013, China

[2]School of Energy and Power Engineering, Huazhong University of Science and Technology, Wuhan 430074, China

[3]Key Laboratory for Thermal Science and Power Engineering of Ministry of Education, Department of Engineering Mechanics, Tsinghua University, Beijing 100084, China

[†]These authors contributed equally to this work.

*Corresponding Emails:

zcliu@hust.edu.cn (Z. C. L.);

maweigang@tsinghua.edu.cn (W. G. M.)

**Abstract**

The jumping-droplet condensation, namely the out-of-plane jumping of condensed droplets upon coalescence, has been a promising technical innovation in the fields of energy harvesting, droplet manipulation, thermal management, *etc.*, yet is limited owing to the challenge of enabling a sustainable and programmable control. Here, we characterized the morphological evolutions and dynamic behaviors of nanoscale condensates on different nanopillar surfaces, and found that there exists an unrevealed domino effect throughout the entire droplet lifecycle and the coalescence is not the only mechanism to access the droplet jumping. The vapor nucleation preferentially occurs in structure intervals, thus the formed liquid embryos incubate and grow in a spatially confined mode, which stores an excess surface energy and simultaneously provides a asymmetric Laplace pressure, stimulating the trapped droplets to undergo a dewetting transition or even a self-jumping, which can be facilitated by the tall and dense nanostructures. Subsequently, the adjacent droplets merge mutually and further trigger more multifarious self-propelled behaviors that are affected by underlying surface nanostructure, including dewetting transition, coalescence-induced jumping and jumping relay. Moreover, an improved energy-based model was developed by considering the nano-physical effects, the theoretical prediction not only extends the coalescence-induced jumping to the nanometer-sized droplets but also correlates the surface nanostructure topology to the jumping velocity. Such a cumulative effect of nucleation-growth-coalescence on the ultimate morphology of droplet may offer a new strategy for designing functional nanostructured surfaces that serve to orientationally manipulate, transport and collect droplets, and motivate surface engineers to achieve the performance ceiling of the jumping-droplet condensation.

1. Introduction

Vapor condensation, as a ubiquitous energy transfer process in nature, has been widely exploited in many industrial contexts, from the energy management [1] to water-energy nexus [2], seawater desalination [3] and power generation [4]. For nearly a century, the pursuit of an energy-efficient condensation never seems to cease [5]. It has been known that dropwise condensation, where the condensate forms discrete droplets, exhibits a heat transfer coefficient that is 4-7 times higher than filmwise condensation [6], due to the considerable thermal resistance of a continuous liquid film adhered to substrate. For dropwise condensation on a smooth hydrophobic surface, heat and mass transfer rely on the condensate removal effect, and droplets are passively shed under gravitational force, whose critical departure size $D_c$ is constrained to the capillary length, $D_c \sim 1$ mm for water [7]. Promisingly, the recent development of micro/nanostructured superhydrophobic surfaces has paved a new pathway to further diminish the droplet-surface affinity [8-10]. Especially in the past decade, a novel mode deriving from the conventional gravity-driven dropwise condensation was discovered, where the merged microdroplets could spontaneously jump away from a surface without any external assistance [11].

Termed "jumping-droplet condensation" enables a reduction in the departing droplet size by several orders of magnitude, and thus has been confirmed to further enhance heat transfer by up to 100% [12,13]. Attracted by its great potential for self-cleaning [14,15], anti-icing [16,17], energy harvesting [18] and electronics cooling [19,20], many follow-up studies have explored the hydrodynamics and energetics of this self-propelled jumping behavior [21-23]. However, there is no consensus on how small a coalesced droplet can detach from a surface [24-29], which is associated with the Ohnesorge number $Oh = \mu/\sqrt{\rho\sigma R}$ that characterizing the relative importance of viscous versus capillary-inertial effects, where $\mu$, $\rho$, $\sigma$ and $R$ are the liquid dynamics viscosity, the liquid density, the surface tension and the radius of droplet, respectively. While earlier researches once suggested that jumping events will not occur below a critical size of $D_c \sim 10$ μm (in large $Oh$ range) [11,24,30], the minimum jumping-droplet radius $R_c$ has exceeded the submicron scale in recent experimental observations [25], even some numerical evidences indicate that nanoscale coalescence-induced jumping should be possible [28,31-34]. It was revealed that $R_c$ is also dependent on the surface topology and the finer structure would enable smaller jumping droplets [27], so some intriguing questions arise accordingly, how to predict the critical departure size for a nanostructured surface and whether the self-removal can be achieved for the naturally grown (condensed) nanodroplets? Unfortunately, to date, these issues have not been exactly solved despite many efforts to investigate the nanometric jumping droplets [28,34-36].

From an energetic standpoint, the surface energy released upon coalescence $\Delta E_S$

is partially converted into the kinetic energy of droplet $E_k$ after overcoming the adhesion work $E_{adh}$ and the viscous dissipation $E_{vis}$ [37,38]. It has been established that viscous effects dominate the coalescence process and suppress the jumping as the droplet size decreases (as $Oh$ increases toward unity), so there is a theoretical critical jumping-droplet radius $R_c$ and a cut-off Ohnesorge number $Oh_c$. However, the existing models generally predict the scaled jumping velocity $U_j^*$ of micrometer-sized droplets ($U_j^* = U_j/U_{ci}$, where $U_{ci} = \sqrt{\sigma/\rho R}$ is the capillary-inertial velocity) [30,39-41], which is uncorrelated with the underlying surface structure and little is known about the coalescence dynamics and energetics at a smaller length scale [29]. In addition, the prediction from continuum physics becomes far from certain with decreasing system size because of the inescapable nano-physical effects [42,43]. Therefore, to rationally extend the prediction of the surface-dependent jumping velocity to the nanoscale and provide a more accurate energy analysis, the surface structure topography and the prominent factors at the nanoscale should be considered comprehensively.

In this study, we characterized the morphology evolutions and dynamic behaviors of nanoscale condensates over the entire lifecycle and elucidate many intriguing phenomena that are neglected in experimental observations, through nonequilibrium molecular dynamics (MD) simulation and energy-based theory analysis. It is found that the condensed droplets sequentially possess various morphologies and self-propelled transitions in nucleation, growth and coalescence stages, which not only diversify the droplet migration and jumping, but also show a cumulative effect on the ultimate wetting state of droplet. Furthermore, we clarified that the spontaneous removal is also possible for the condensed nanodroplets and the jumping-droplet condensation is not actually a unitary pattern only dominated by the coalescence mechanism. These findings can provide a theoretical foundation to rationally design nanostructure surfaces for self-cleaning, droplet manipulation and water collection.

## 2. Simulation Methods

To delve deeper into the morphological evolutions of nanoscale condensates during condensation, particularly the incipient behaviors of nucleating embryos, large-scale MD simulations were conducted as depicted in Figure 1. In the model system with the dimensions of 54.3 nm × 54.3 nm × 54.3 nm, the nanopillar surfaces with various topologies and wettability (see Table S1 in Supplementary Material (SM) for detailed geometric dimensions, including the width $W$, spacing $S$ and height $H$ of the nanopillars, the roughness and the solid fraction of surface) were initially submerged in the saturated water vapor, and the bottom substrate is divided into the fixed layer, the phantom layer and the conduction layer [44]. For convenience, all geometric dimensions are normalized by $W$ (3.1 nm), and the dimensionless parameters are presented with an asterisk, e.g., $S^* = S/W$, $H^* = H/W$. To provide a stable temperature and pressure control in the saturated state, as shown in Figures S1 and S2 of SM, the top quarter of vapor domain was set as a vapor/heat source to continuously supply the saturated vapor, through the GCMD method that is a combination of the grand canonical Monte Carlo (GCMC) and the MD simulation [45,46].

To reduce the computational cost, the simpler copper atoms (Cu) were chosen to construct the solid surface with tunable wettability, and a coarse-grained (CG) model [47], where each CG water bead (W) represents four water molecules, was adopted to constitute the vapor based on the saturation density at a given temperature. Morse potential was employed to describe the water properties exactly, and the interactions of Cu-Cu and Cu-W were respectively described by the embedded atom model (EAM) potential and the 12-6 Lennard-Jones (LJ) potential, the forcefields parameters are listed in Table S2 of SM [48]. To quantitatively evaluate the wetting capacities of these solid surfaces, we calculated the intrinsic contact angles $\theta_0$ of sessile cylindrical nanodroplets, eliminating the size-dependent deviation caused by line tension (see section S2 in SM for details) [49,50].

All these simulations involve two stages and were performed in LAMMPS software package [51]. To develop a steady solid-vapor coexistence at a temperature of 500 K, the entire system was first pre-equilibrated in a canonical (NVT) ensemble with the Nose-Hoover method for 1 ns. Then in the main simulation stage of 10 ns, the upper GCMD region was switched to a grand canonical (μVT) ensemble, while the remaining parts were integrated in a microcanonical (NVE) ensemble with a Langevin thermostat controlling the temperature of substrate (phantom layer) at 300 K (see section S1 in SM for more simulation details).

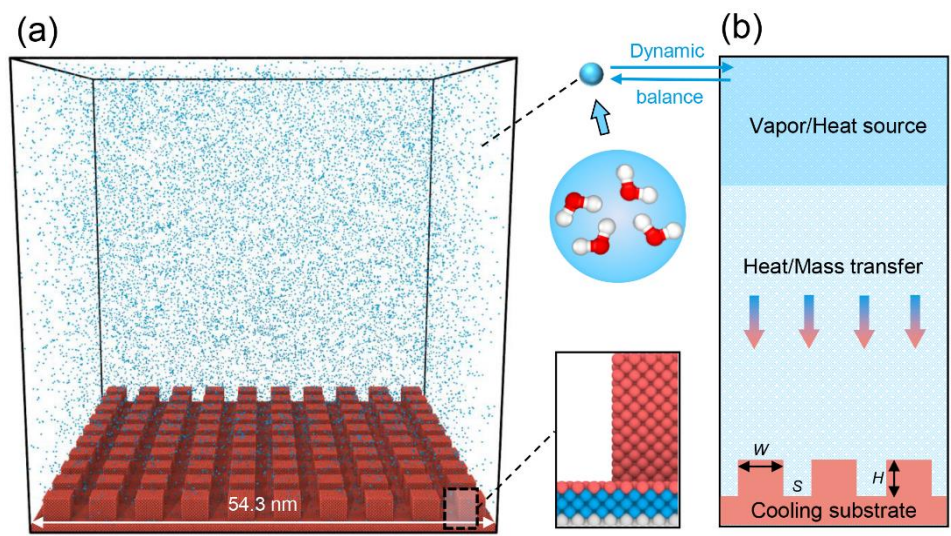

Figure 1. (a) Initial configuration of the large-scale MD condensation model with a side-length of 54.3 nm. The bottom substrate is divided into the fixed layer, the phantom layer and the conduction layer, and the saturated vapor consists of the 4-site coarse-grained water beads. (b) Schematic illustration of the simulation domain settings. The top quarter region is used to replenish the vapor and heat by GCMD methods, and the water beads from the high-temperature vapor phase move to the cooling substrate and condense continuously, coupling with heat and mass transfer. The geometric parameters $W$, $S$ and $H$ denote the width, spacing and height of nanopillars, respectively.

### 3. Results and discussions

Based on the simulation results, the dynamic evolutions of condensed droplets were analyzed and discussed in the order of droplet lifecycle, and the condensed droplets forming on each surface were observed to range from ~1 nm to ~20 nm under our simulation conditions.

*3.1 Preferential Nucleation*

In general, vapor condensation originates from the formation of liquid nuclei. To probe the nucleation characteristics, the nucleating embryos at the substrate-vapor interface were visualized. As shown in Figure 2a, accompanied by the collisions with low-temperature surface, initially dispersed vapor beads were gradually captured and aggregated into clusters, *i.e.*, the condensing nuclei. Significantly, most nuclei regularly distributed within the surface roughness rather than atop the structure, showing a spatial preference of nucleation sites (similar to the nucleation characteristics of other simulated surfaces, see Video S1 of SM). This suggests that, for a homogeneous surface, liquid embryos nucleate in the crevices inevitably and the resultant droplets ought to immerse the surface if current trend continues, however, it is obviously inconsistent with the diverse wetting states (*e.g.*, suspended Cassie, immersed Wenzel and partial wetting) as observed in reality. We therefore infer that the ultimate morphology of a condensed droplet may be determined collaboratively by each stage of the droplet lifetime, and these cumulative effects will be further discussed in subsequent sections.

To understand the mechanism of preferential nucleation from a molecular-level view, we calculated the potential energy $U$ of a water CG bead at the solid-liquid interface, which represents an energy restriction. Figure 2b shows the distribution of $U$ at a horizonal (XY-plane) cross section, where there exists a distinct spatial nonuniformity that can affect vapor nucleation behaviors, and the low-potential-energy regions (nanostructure bases) approximately match with the initial nucleation sites. In these regions, the potent solid-liquid interactions promote the exchange of thermal energy, and meanwhile, the water beads are restricted more strongly. As a result, the vapor is inclined to nucleate nearby the nanostructure base, and the generated liquid embryos further incubate the condensate nanodroplets within a single unit cell of roughness (nanostructure valley), exhibiting a spatially preferential nucleation. Furthermore, to corroborate this view, the Gibbs free energy barrier of nucleation was analyzed according to the classical nucleation theory (CNT) [52].

For heterogeneous nucleation, provided that the radius of a liquid embryo is the critical nucleus size $r_e$, then its nucleation energy barrier $\Delta G$ can be expressed as [53,54]

$$\Delta G = -\frac{\rho V}{M}k_B N_A T_s \cdot \ln s + \sigma_{LV}(A_{LV} - A_{SL} \cdot \cos\theta) \qquad (1)$$

where $\rho$, $V$ and $M$ are the liquid density, the volume of embryo and the molar mass of liquid. $k_B$, $N_A$ and $T_s$ represent the Boltzmann constant, the Avogadro number and the surface temperature. $s$ is the supersaturation defined as the ratio of vapor pressure $P_\infty$ to the saturated vapor pressure $P_s$ at $T_s$. $\sigma_{LV}$, $A_{LV}$, $A_{SL}$ and $\theta$ represent the surface tension of liquid, the liquid-vapor interfacial area, the solid-liquid interfacial area and the Young's contact angle. Obviously, $\Delta G$ is directly connected with the nucleus shape that affected by the surface topology features, under a given thermodynamics condition. There are two typical nucleation configurations for the pillared surface studied here, namely the spherical-cap nucleus on the pillar top and the spherical-wedge nucleus at the pillar base, whose nucleation energy barriers $\Delta G_1$ and $\Delta G_2$ were calculated based on the above CNT model, as illustrated in Figure 2c (calculation details of the nucleation energy barrier refer to the Section S3 in SI). The liquid-vapor boundary of nuclei embedded in the cavity is smaller than that atop the structure, and the reduced interfacial free energy lowers the energy barrier to overcome ($\Delta G_2 < \Delta G_1$). Hence, nucleation is more readily to occur in structure intervals. Moreover, it is found that $\Delta G_2$ is also dependent on the structure dimensions. As the spacing $S$ shrinks, $i.e.$, when $S < (1 - \cos\theta)r_e$, the nucleus becomes sandwiched between the sidewalls of pillars, so $\Delta G_2$ is further lowered due to the reduction of liquid-vapor interfacial area.

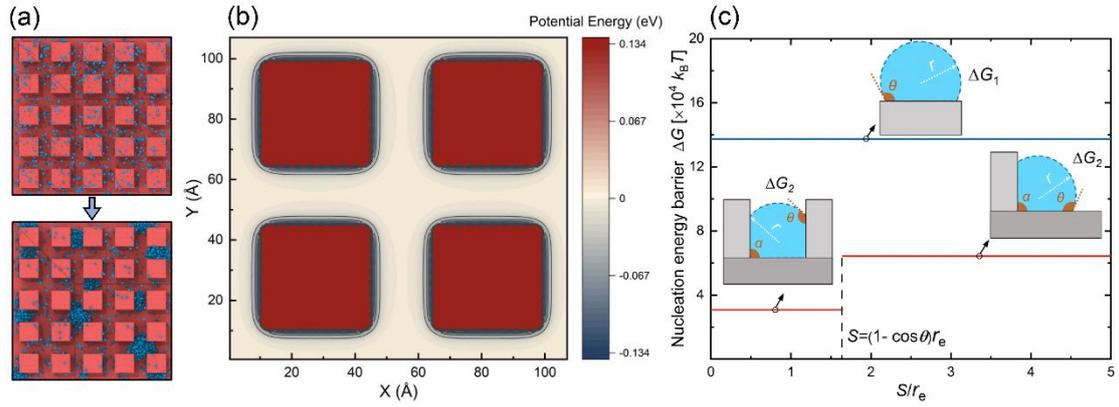

Figure 2. (a) Snapshots of the heterogeneous nucleation for water. (b) Potential energy distribution of a CG water bead at a horizonal (XY-plane) cross section of the unit cell (2×2) of nanopillar array. (c) Nucleation energy barriers at different positions of the unit cell of nanopillar array.

*3.2 Spatially confined growth*

After preferentially nucleating within structures, the liquid nuclei grow inside the unit cell of nanopillars continuously, and there emerged three representative modes classified by the dynamic behaviors of the generated nanodroplets (see Video S2 in SM). As shown in Figure 3a, in the incipient growth stage, every nucleated droplet exhibits a roughly spherical-cap shape, but become deformed immediately as the growing droplet touches the sidewalls of nanopillars. After filling the footprint of the unit cell, the confined droplet grows upward until its top interface exceeds the upper edge of pillars, and then the bulging droplet head (bulb) starts to expand both laterally and vertically, free from the constraint of the structures. But unexpectedly, the further growth of the droplet bulb leads to various state transitions, the droplet body might still be trapped inside structure valley (impaled Wenzel state, mode I), also might be pumped to structure tops (suspended Cassie state/dewetting transition, mode II) or even be ejected from the surface (self-jumping, mode III). We have to emphasize that the droplet grows separately without any disturbance; moreover, the status of the growing droplet on each surface is not exactly the same. So it is reasonable to conclude that these different outcomes are relevant with the surface structure properties.

Figure 3b schematically shows the above spatially-confined growth process, whose dynamic evolution is also visualized by the time-lapse snapshots in Figure 3d. In the early incubation period with the droplet being fully embedded in the pillar valley, upward growth is more energetically favorable *versus* sideways inflation [27], and the droplet shape could be approximated as an axisymmetric cylinder bounded by the rectangular pillars at the corners. Once the longitudinal triple-phase contact line reaches the height of pillars, the curvature radius of top interface $R_1$ increases with the inflating droplet bulb whose base is pinned at the pillar top edges, while the curvature radius of bottom interface $R_2$ still remains unchanged, that will generate an upward Laplace pressure difference $\Delta P$ [55-57]. When the droplet bulb grows to a certain size, the driving force is large enough to overcome the surface adhesion force, and further triggers the detachment of the squeezed droplet body from structure valleys.

To corroborate the above force analysis, the curvature radius $R_1$ and the dynamic contact angle $\theta$ (as marked in Figure 3b) were extracted from the droplet morphology, as shown in Figure 3c, where both $R_1$ and $\theta$ first fluctuate in a certain range and turn to rise rapidly. In the incubation stage, the value of $R_1$ (17.78 Å) obtained from the fitting data is roughly equal to the calculated value of $R_2$ (16.84 Å) estimated as $S/2|\cos\theta_0|$, but this balance is soon disturbed in the later burst stage. As the growing droplet emerges from the valley, $R_1$ increases dramatically with the fast expansion of the droplet bulb, in contrast, $R_2$ stays almost constant under the shape restriction from structures. As a result, the steady increase in curvature difference leads to an ever-increasing Laplace pressure difference $\Delta P$ ( $\Delta P = P_2 - P_1 \approx 2\sigma(1/R_2 - 1/R_1)$, given the curvature difference in the XZ-plane is approximately equal to that in the YZ-plane). Furthermore, from the visualized droplet shape and the variation tendency of $\theta$, it is observed that the droplet bulb grows in a constant contact radius (CCR) and a constant contact angle (CCA) mode successively, which can be elucidated by a local force analysis [58,59]. Since the contact line of the droplet bulb is initially confined to the pillar edges under the pinning force $F_P$, the droplet growth compels the dynamic contact angle $\theta$ to rise firstly (the contact radius $R_c$ remains constant), and thus provide an increasingly powerful depinning force $F_D = 2R_c\sigma(\cos\theta - \cos\theta_0)$. As $\theta$ increases to a critical value, i.e., the advancing contact angle $\theta_A$, $F_D$ can overcome $F_P$ to move the contact line, resulting in a both lateral and vertical expansion of the droplet bulb with a roughly unchanged $\theta$.

To investigate the dynamics of the growth-induced self-jumping, the height $Z$ and the vertical velocity $U_Z$ of centroid were extracted from the droplet trajectory. As shown in Figure 3d, both of them remain constant initially and then increase in time with a power-law relation. In the burst period, the rapid retraction of the droplet tail causes a sharp increase in $Z$ and $U_Z$, and due to the pinning effects on the droplet tail that is initially adhered to the structure valley floor, the vertical velocity of the droplet tail $U_t$ is significantly higher than that of the droplet bulb $U_b$ (see Figure S9 in SM).

The pinning force hinders the detachment of droplet tail and further stretches the droplet body, until the growing droplet bulb attains a sufficient Laplace pressure difference. Then, the abrupt depinning of the droplet tail triggers an explosive retraction of the originally elongated droplet body, and the consequent upward flow and momentum propel the droplet bulb to move.

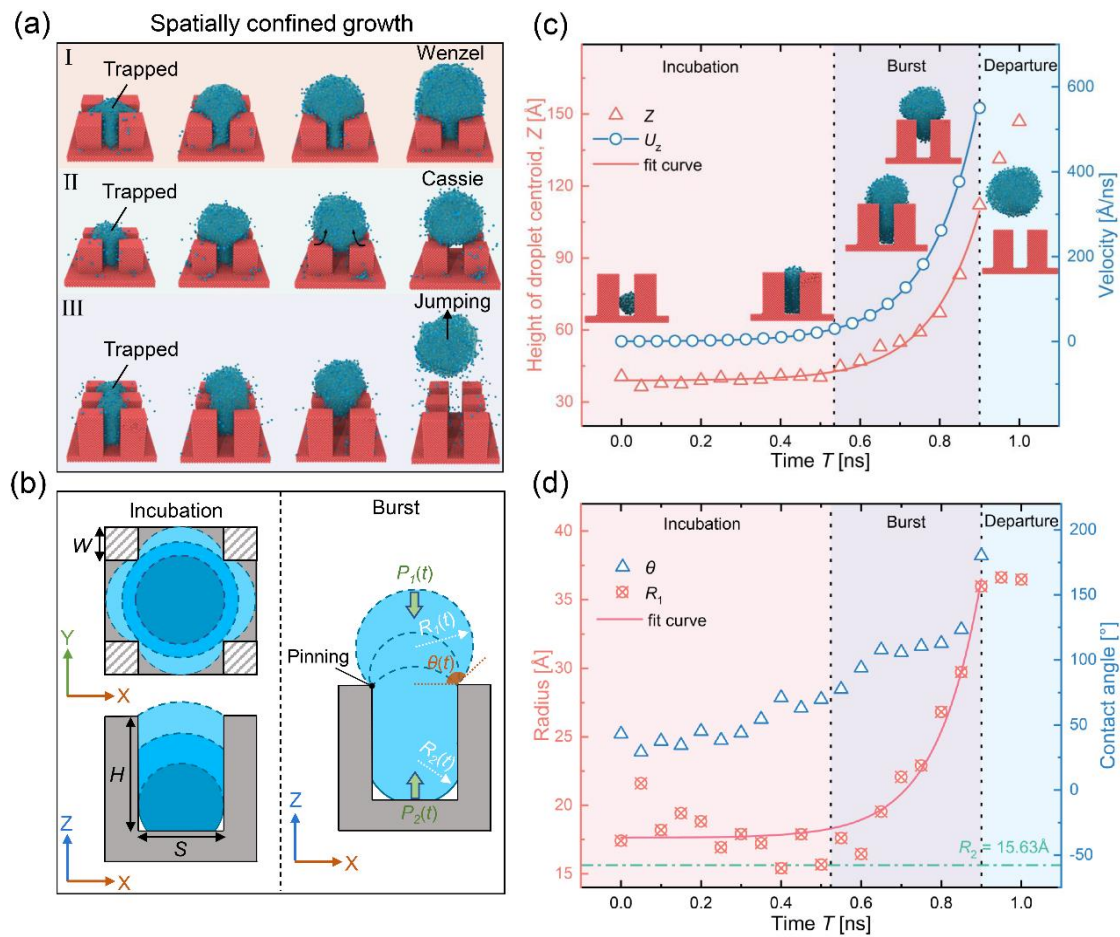

Figure 3. (a) Spatially-confined growth modes of an isolated droplet on different nanopillar surfaces: growth-induced dewetting transition and self-jumping (without any coalescence events). (b) Schematic of the growth process, the squeezed droplet grows upwards firstly until it emerges from the pillars valley (the incubation stage), and followed by the rapid expansion of droplet head (the burst stage). (c) Temporal

evolutions of the apparent dynamic contact angle $\theta$ and the curvature radius $R_1$ at the upper liquid-vapor interface of the growing droplet, the curvature radius $R_2$ at the bottom interface is estimated to be $S/2|\cos\theta_0|$. (d) Droplet centroid height $Z$ and vertical velocity $U_Z$ as a function of time. Insets illustrate the typical position and morphology of the growing droplet in each corresponding stage.

Figure 4a is a phase diagram of the three growth modes. *i.e.*, Wenzel, Cassie and self-jumping, as a function of the pillar topography (the dimensionless spacing $S^*$ ($S/W$) and height $H^*$ ($H/W$)) and wettability (the intrinsic contact angle $\theta_0$) based on all the numerical simulation cases. From this figure, one can know how to design the surface roughness to effectively control the wetting state of a growing droplet. The droplet undergoes Wenzel state, Cassie state and self-jumping state successively, with the decrease of $S^*$ or with the increase of $H^*$ and $\theta_0$, so we can deduce that the tall and dense nanostructures with a good hydrophobic performance could intensify the mobility of droplet, which is attributed to the competition between the excessive surface energy stored in the deformed droplet and the frictional dissipation at the interface. Additionally, to provide an energic insight into these diverse wetting states, we conduct an energy analysis on the spatially-confined droplet growth.

In terms of energy arguments, the deformed droplet possesses a larger surface area compared to the spherical droplet with an identical volume. It follows that an excessive surface energy $\Delta E_s$ is stored in the squeezed droplet caused by the spatial confinement of the pillar sidewalls, as a result, the spherical shape is energetically favorable and the released $\Delta E_s$ might be partially converted into the kinetic energy during the shape recovery . Based on the above analysis, we obtained the $\Delta E_s$ ($\Delta E_s = E_{\text{initial}} - E_{\text{final}}$) and the scaled excessive surface energy $\Delta E_s^*$ ($\Delta E_s/E_{\text{initial}}$) by calculating the surface areas and the droplet volume (Section S4, Supplementary Material), to evaluate the potential for droplet dewetting transition. As shown in Figures S7 and S8 in SM, our model first considers a droplet that is constrained inside a unit cell just prior to state transition ($E_{\text{initial}}$), whose critical radius $R_c$ is determined by the pillar topography and wettability [60], and the deformed droplet is approximately a combination of a spherical-cap head with a volume of $V_1$ and an incomplete cylindrical body with a volume of $V_2$. Then, supposing this droplet recover to a sphere and jump away from the surface ($E_{\text{final}}$), its final radius $R_{\text{final}}$ can be obtained from the volume conservation $V_1 + V_2 = 4\pi R_{\text{final}}^3/3$. Taking the intrinsic contact angle $\theta_0$ as 120°, we draw the theoretical results of $\Delta E_s^*$ as a function with the dimensionless pillar spacing $S^*$ and height $H^*$ (Figure 4b). The result suggests that $\Delta E_s^*$ increases with increasing $H^*$ or decreasing $S^*$, that is, sufficiently tall and dense nanopillar serve to improve the excessive surface energy and further prompt the trapped droplet to change its wetting state, which is consistent with the aforementioned simulation results.

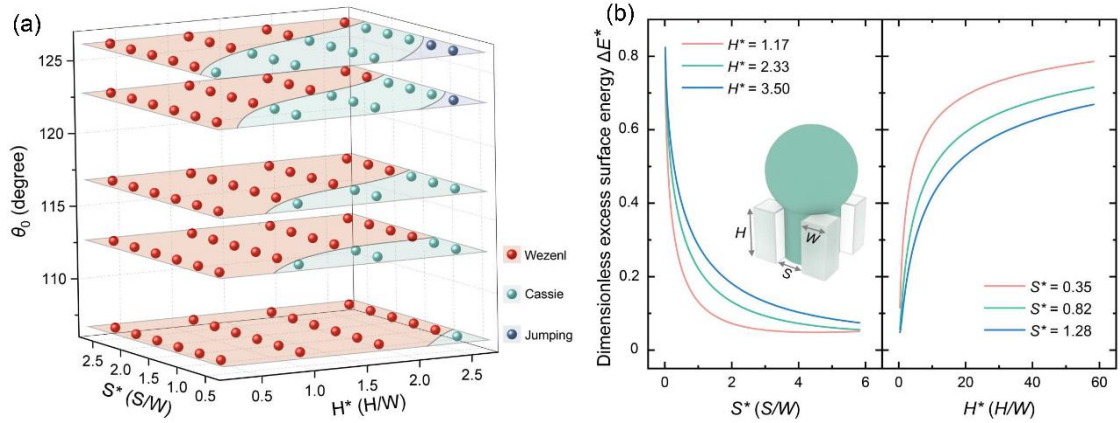

Figure 4. Different droplet growth modes and the energetic analysis. (a) Three-dimensional regime map of spatially-confined droplet growth as a function of the surface structure topography ($S^*$, $H^*$) and wettability ($\theta_0$). (b) Relations between the scaled excess surface energy $\Delta E_s^*$ ($\Delta E_s/E_{\text{initial}}$) and the dimensionless geometric parameters ($S^*$, $H^*$) of rectangular pillar ($\theta_0 = 120°$).

*3.3 Coalescence-induced transition*

Following the spatially-confined growth with three modes, the two adjacent droplets merge with each other, thereby leading to a more diverse set of outcomes on different nanopillar surfaces, as shown in Figure 5. Noted that they are rarely noticed and discussed in previous experimental evidence, and we classify them into four regimes according to the initial state of droplet. Since the nucleated droplets tend to grow within structures, the coalescence events usually start between two small Wenzel droplets (mode I), as shown in Figure 5a, and the newly formed droplets present three distinct wetting states (Wenzel, Cassie or jumping) under different solid-liquid interface interactions (see Video S3 in SM). In addition, the small Wenzel droplet may also merge with a slightly larger Cassie droplet (mode II) that derived from the growth- or coalescence-induced state transition of a small Wenzel droplet, and further transform into a large Cassie droplet or even depart from the surface (see Video S4 in SM). Soon afterwards, the resultant flying droplet might touch and assimilate other sessile Wenzel or Cassie droplets around its flight path (mode III), and stimulate them to change their status again or jump out-of-plane (see Video S5 in SM). Finally, it comes to the coalescence of two Cassie droplets (mode IV) that has been often observed in previous studies, the coalesced large droplet might remain its original wetting state or will probably lead to a jumping as well (see Video S6 in SM). Furthermore, the coalescence-induced transition also occurs among the multiple droplets with various states (see Figure S14 and Video S7 in SM).

The above findings complement the experimental observations and improve the understanding for the droplet morphology evolutions, indicating that the jumping-droplet condensation actually consists of multifarious state-transition and self-jumping events occurring among various droplets, rather than as a unitary pattern dominated by the coalescence-induced jumping of Cassie droplets. Moreover, it is interesting to note that the final wetting state of a condensed droplet seems more like a chain-reaction result of these sequential nucleation-growth-coalescence events, and our simulations suggest that the nanopillar surface with low solid fraction and strong hydrophobicity generally endows the condensed droplets with a good mobility (suspended Cassie state or coalescence-induced jumping), and the reason will be discussed later in this study.

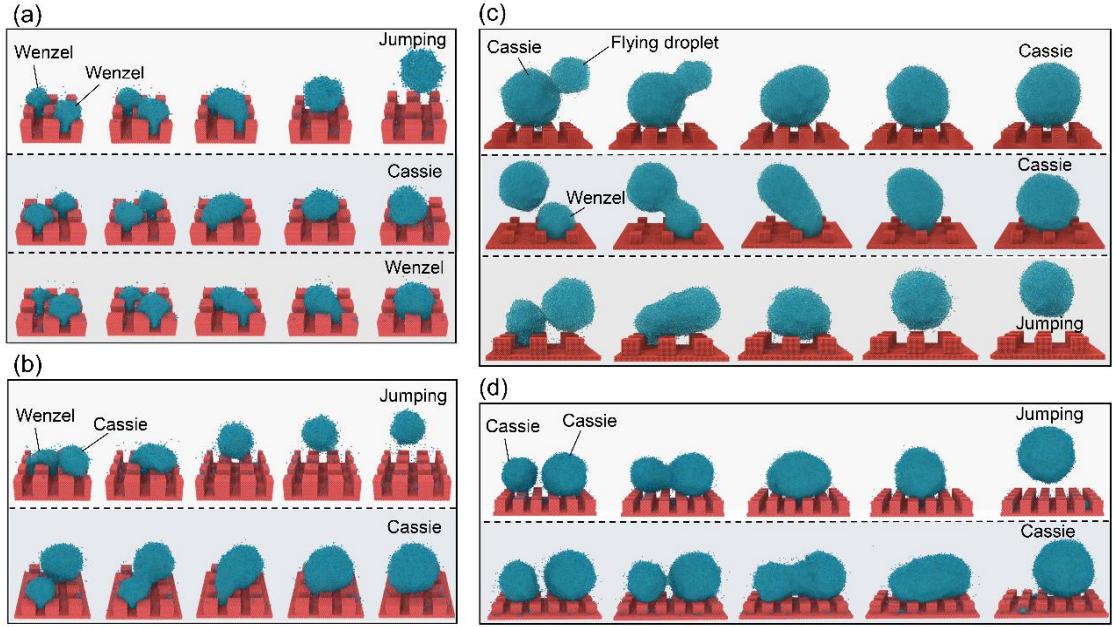

Figure 5. Comparison of the time-lapse snapshots for the condensed droplet coalescence modes on different nanopillar surfaces. All of these two-droplets coalescence events may potentially cause a spontaneous jumping or a dewetting transition. (a) Mode I, Wenzel-Wenzel droplets coalescence. (b) Mode II, Wenzel-Cassie droplets coalescence. (c) Mode III, coalescence between a flying droplet and a sessile droplet. (d) Mode IV, Cassie-Cassie droplets coalescence.

The jumping case in Figure 5d was selected as an example to analyze the dynamics characteristics during droplet coalescence. Based on the position coordinate of the droplet centroid, we calculated the vertical velocity of the coalesced droplet $U_Z$ and the horizontal radial velocity of each droplet, $U_{X1}$ and $U_{X2}$. Figures 6a and 6b respectively show their temporal evolutions, which could be divided into four phases according to the corresponding droplet morphology visualized in the insets. Start from the formation and expansion of the liquid bridge (0 - 0.1 ns), the centroid height is lowered ($U_Z < 0$) as the liquid mass migrates toward the center ($U_{X1} > 0$, $U_{X2} < 0$) driven by the Laplace pressure. When the downward-moving liquid bridge contacts and impacts the substrate, the coalesced droplet accelerates upwards until $U_Z$ reaches the maximum value of 0.17 Å/ps at 0.25 ns, and the lateral retraction slows down simultaneously (both $U_{X1}$ and $U_{X2}$ decrease). During the subsequent droplet detachment, $U_Z$ decelerates to a critical value ~ 0.08 Å/ps (the jumping velocity $U$) at 0.45 ns under the adhesion force exerted by surface. Meanwhile, the direction of $U_{X1}$ and $U_{X2}$ are inverted symmetrically, accompanied by a periodic oscillation of the coalesced droplet. At last (> 0.45 ns), the horizontal radial velocities attenuate to zero, and the kinetic energy of departing droplet is dissipated gradually due to the resistance from vapor particles and clusters. It should be noted that the variations of $U_{X1}$ and $U_{X2}$ show a reversed-phase oscillation in general, but the velocity amplitudes are inconsistent because the mass of each coalescing droplet is different (the diameters $D_1 \approx 13$ nm, $D_2 \approx 15$ nm).

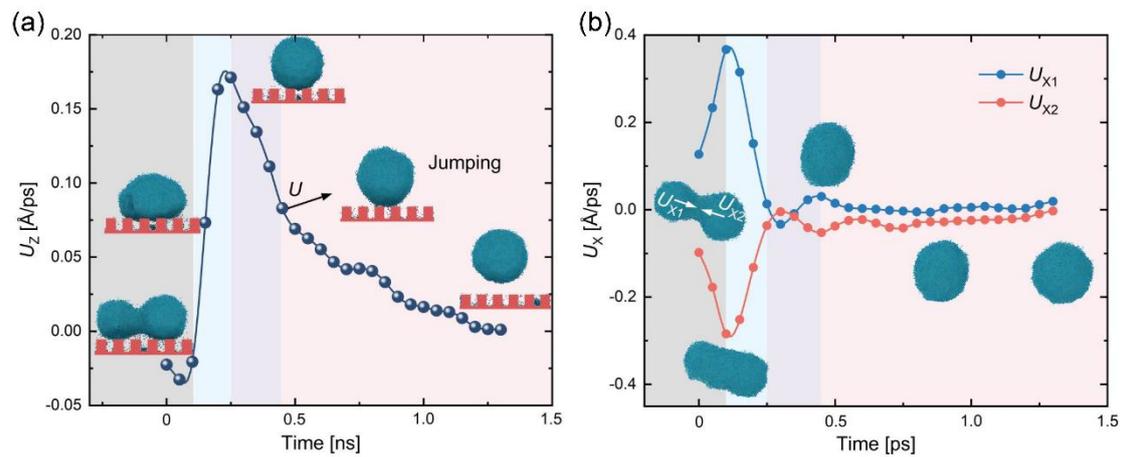

Figure 6. Dynamic analysis on the coalescence-induced jumping of condensed nanodroplets. (a) Temporal variation of the vertical centroid velocity $U_Z$. The series of snapshots show the typical droplet morphology in these four specific stages (front view). (b) Temporal variation of the horizontal radial velocity for each droplet, $U_{X1}$ and $U_{X2}$.

Insets visualize the morphology evolution of the coalesced droplet (top view).

Generally, the droplets coalescence on a surface follows the energy-conservation principle, however, the existing models developed from the basic energy-balance equation, $\Delta E_s = E_{adh} + E_{vis} + E_k$, mostly treat the surface underneath droplets as smooth because of its relatively small feature size of roughness [24,61-64]. Hence, they overlook the role of the underlying surface structure, which will be more prominent as the droplet size descends. Furthermore, the above visualization results indicate that the state transitions induced from coalescence are varied with the structure topography and wettability, so we propose a modified model to rationalize this relationship and extend the prediction of the jumping velocity to the nanometric length scale. Given the coalescence process of two same-sized droplets sitting on the nanopillar arrays, as schematically shown in Figure S10 of SI, the initial total surface energy could be evaluated as

$$E_S = 2[2\pi(1 - \cos\theta)R^2\sigma_{LV} + \pi R^2\sin^2\theta(1 - \varphi_s)\sigma_{LV} + \pi R^2\sin^2\theta\varphi_s\sigma_{SL}] \quad (2)$$

where the apparent contact angle $\theta$ can be obtained by the Cassie-Baxter equation $\cos\theta = \varphi_s(\cos\theta_0 + 1) - 1$ and $\theta_0$ is the intrinsic contact angle, $\varphi_s = W^2/(W + S)^2 = 1/(1 + S^*)^2$ is the surface solid fraction, $R$ is the radius of each droplet, $\sigma_{LV}$ and $\sigma_{SL}$ are the liquid-vapor and solid-liquid surface tensions, respectively. By contrast, the total surface energy after coalescence-induced jumping could be calculated as $E_S' = 4\pi(R')^2\sigma_{LV} + \pi R^2\sin^2\theta\varphi_s\sigma_{SV}$, where the radius of the coalesced droplet $R' = \sqrt[3]{\frac{2-3\cos\theta+\cos^3\theta}{2}}R$ is calculated on the basis of the volume conversation, then we can obtain the available excessive surface energy $\Delta E_S = E_S - E_S'$, The surface adhesion work $E_{adh}$ could be expressed by Young-Dupre equation

$$E_{adh} = 2\pi R^2 \sin^2\theta (1 + \cos\theta)\varphi_s \sigma_{LV} \tag{3}$$

The viscous dissipation $E_{vis}$ is mainly incurred by the elaborated internal flow in the droplet (fluid convection or diffusion), and could be estimated as[24]

$$E_{vis} = 2\int_0^{\tau_{vis}} \int_V \Phi dV dt \approx 2\Phi V \tau_{vis} \tag{4}$$

where $V = \pi(2 - 3\cos\theta + \cos^3\theta)R^3/3$ is the volume of each droplet, $\tau_{vis}$ is the viscous dissipation time, and $\Phi$ is the dissipation function. It has been confirmed that the viscous dissipation during coalescence is dominated by the droplet lateral motion, and thus the dissipation function $\Phi$ is estimated to be

$$\Phi \approx \frac{1}{2}\mu(\frac{U_x}{R})^2 \tag{5}$$

where $\mu$ is the liquid dynamic viscosity, $U_x$ is the average merging velocity in the lateral direction, which could be estimated as[65]

$$U_x \approx 2(\theta + \sin\theta\cos\theta)\sigma_{LV} R \cdot \tau_{vis}/\rho V \tag{6}$$

for the convenience of calculation, the dissipation time $\tau_{vis}$ is approximated by the coalescence time $\tau$, which is given by[25]

$$\tau = \frac{\mu r_c}{\sigma_{LV}} + (\frac{\rho}{\sigma_{LV} D_0^4})^{\frac{1}{2}} \cdot (R^{\frac{3}{2}} - r_c^{\frac{3}{2}}) \tag{7}$$

where $\rho$ is the liquid density, the constant $D_0$ is taken as 1.39, and $r_c \approx 8\mu\sqrt{R/(\rho \sigma_{LV}^3 D_0^4)}$ is the critical bridge radius at the viscous-to-inertial regime transition moment. As such, the viscous dissipation $E_{vis}$ can be obtained by substituting Eq. 5-7 into Eq. 4.

However, the basic energy-balance equation mentioned above is incompetent to exactly describe the coalescence of nanodroplets, thus some extra nano-physical effects need to be considered in the energy analysis. As the droplet size approaches to nanoscale, the energy dissipation created by contact-line movement $E_{cl}$ is increasingly prominent, which is determined by the contact-line friction coefficient $\mu_{cl}$, the characteristic contact-line velocity $u$, the characteristic length $L$ and the contact-line dissipation time $\tau_{cl}$, given by[66]

$$E_{cl} \approx \mu_{cl} u^2 L \tau_{cl} \tag{8}$$

where $\mu_{cl}$ could be approximated by $\mu$, $u$ is on the same order of magnitude as $U_x$, $L$ is defined as distance between the two initial droplet-surface contact points, here, $L \sim 2R$, and $\tau_{cl}$ could be approximated by $\tau$. Moreover, the intrinsic line tension effect also becomes unneglectable with the shrink of spatial scale, and the energy dissipation $E_{lt}$ resulting from line tension could be evaluated as[43]

$$E_{lt} = \sigma_\kappa L_{lt} \tag{9}$$

where $\sigma_\kappa$ is the line tension coefficient and its value is taken as $2 \times 10^{-10}$ N [49,50], $L_{lt}$ is the total contact line length, which is contributed by the internal contact line $L_{in} = \frac{4\pi}{W}\left(\frac{R\sin\theta}{1+S^*}\right)^2$ and the contact line boundary $L_{ex} \approx 1.738\pi \frac{R\sin\theta}{(1+S^*)^2} \cdot (2.184W + 0.288\pi)$. In addition, it should be stressed that the total kinetic energy $E_k$ of coalesced droplet contains the translational part $E_{k,tr}$ and the oscillatory part $E_{k,os}$, while the jumping behavior is power by the $E_{k,tr} = \frac{mU_j^2}{2}$, where $m$ is the mass of coalesced droplet and $U_j$ is the jumping velocity, and the proportion of $E_{k,tr}$ in $E_k$ is given by a piecewise function $g(Oh)$ of the Ohnesorge number $Oh = \mu/\sqrt{\rho\sigma_{LV}R}$ [29]. Finally, combining all of these energy items yields a modified energy-based model (see

fully detailed derivation in Section S5 of SM)

$$E_{k,tr} = g(Oh)(\Delta E_s - E_{adh} - E_{vis} - E_{cl} - E_{lt}) \qquad (10)$$

which formulates the dependence of the jumping velocity $U_j$ on the surface structure topology ($W$, $S$) and wettability ($\theta$) and the droplet size $R$ (or the $Oh$ number).

Figure 7a shows the variation of the inertial-capillary scaled jumping velocity $U_j^*$ ($U_j^* = U_j/U_{ci}$, $U_{ci} = \sqrt{\sigma/\rho R}$) with the $Oh$ number by the present predictions, for a given nanopillar surface with $S^* = 0.875$, $H^* = 1.375$ and $\theta_0 = 120°$ ($\theta \approx 150°$). Compared with two previous theoretical models where the viscous dissipation $E_{vis}$ is given by $E_{vis} = f(Oh)E_r$ ($E_r = \sigma_{LV}R^3$ is a reference energy, the $f(Oh)$ is respectively a power law formula $6.511Oh^{0.273}$ [29] and a linear form $3\pi Oh$ [41]), our present model predicts a wider range of $Oh$ for the possible coalescence-induced jumping, that is, the minimum jumping-droplet radius $R_c$ could reach the nanoscale (about tens of nanometers, see Figure S12 in SM), which basically agrees with our MD results and other numerical simulations. Furthermore, the predicted $U_j^*$ first increases rapidly, and then declines slowly and becomes relatively stable as $Oh$ decreases, which is quite distinct from the monotonic increase tendency commonly assumed by most prior theoretical models. This is attributed to the regulatory effects of surface structure that hasn't been considered before, as shown in Figure S13 of SM, with the decrease of the dimensionless pillar spacing $S^*$, the nanostructure arrays look increasingly like a smooth surface, and consequently the complicated $U_j^* - Oh$ curve gradually degenerates into a simple monotonic curve that can be regarded equivalent to these classical theoretical models. Moreover, it is evident from Figure S12 that the scaled jumping velocity $U_j^*$ increases with increasing dimensionless pillar spacing $S^*$, which implies that the structured surface with a lower solid fraction leads to a more efficient surface-to-kinetic energy conversation. Figure 7b shows the relations of $Oh_C$ with the surface apparent contact angle $\theta$ and the dimensionless pillar spacing $S^*$. Here, $Oh_C$ is the critical (cut-off) Ohnesorge number corresponding to the critical (minimum) jumping-droplet radius $R_c$, and can be used to measure the difficulty levels of the coalescence-induced jumping on different surfaces. It is seen that $Oh_C$ is approximately proportional to $\theta$, and also follows a positive power-law relationship with $S^*$, which theoretically validates that the coalescence-induced jumping is highly dependent on the topology and wettability of surface nanostructure. Also, it suggests that rationally designed slender nanostructures with a good hydrophobicity and a low solid fraction would facilitate the departure of smaller droplets and even enable nanometric jumping.

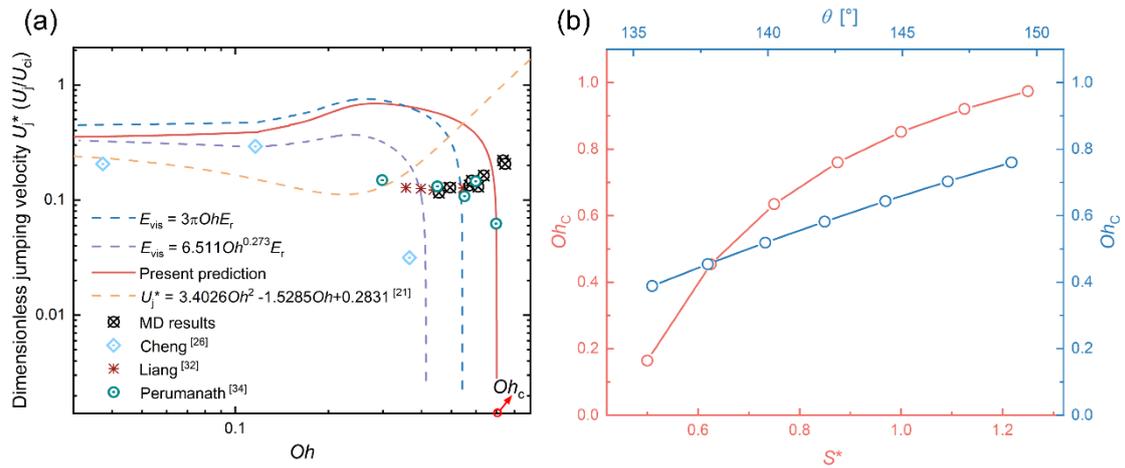

Figure 7. Theoretical prediction for the jumping velocity of coalesced droplet in the nanometer to micrometer size. (a) Variations of the scaled jumping velocity $U_j^*$ with the $Oh$ number based on the present predictions and MD simulations, previous

theoretical predictions, numerical simulations and experimental measurements. Here, $Oh_\text{C} = \mu/\sqrt{\rho\sigma_\text{LV}R_\text{C}}$ is the cut-off $Oh$ number corresponding to the minimum jumping size $R_\text{C}$. (b) Dependence of the $Oh_\text{C}$ on the surface structure topography ($S^*$) and wettability (apparent contact angle $\theta$).

## 4. Conclusions

In summary, we have studied the various self-propelled state-transitions and motions of nanoscale condensates on nanopillar surfaces, by MD simulations and theory analysis. It is found that the nucleating embryos potentially evolve into the droplets with diverse morphologies, showing a cumulative effect of all stages of the droplet lifecycle on the final wetting state. Due to the relatively low nucleation energy barrier, the liquid embryos prefer to incubate in the nanostructures and further grow up with their lateral expansion are restricted by the pillar sidewalls. The resultant squeezed droplets might behave the state transitions (Wenzel-Cassie dewetting transition or even self-jumping) under an asymmetric Laplace pressure that is affected by the nanopillar topography, and the energy analysis reveals that tall and dense structures serve to store more excessive surface energy to facilitate the state transitions. And then, the multifarious coalescence events between these droplets lead to more diverse outcomes and probably trigger the jumping as well, indicating the final morphology of a condensed droplet is a chain-reaction result of these sequential nucleation-growth-coalescence events. By proposing a modified energy-based model considering the nano-physical effects, we extend the prediction of jumping velocity to the nanometric length scale and find the slender nanostructures arrays with a low solid fraction could prompt the coalescence-induced jumping. This work reveals the insightful physical mechanisms of the micromorphological evolutions of a condensed droplet over its entire lifecycle, and suggest that the jumping-droplet condensation is not actually a unitary pattern dominated by the coalescence-induced jumping. We hope that these findings could provide guidance for the surface nanofabrication to achieve the droplet rapid generation, directional transport and efficient removal while also promoting droplet manipulation techniques and self-cleaning surface technology.

**Acknowledgement**

The authors thank the National Supercomputer Center in Tianjin for providing computing resources. This study was supported by the National Natural Science Foundation of China (Nos. 52076088, 52176078 and 52276067) and sponsored by Tsinghua-Toyota Joint Research Fund.

**Supplementary Material**

Simulation details including physical models, potential functions and control methods (Section S1); calculation and characterization of the intrinsic contact angle (Section S2); calculation of nucleation energy barrier (Section S3); Energy calculation of the growth-induced transition/departure (Section S4); Energy analysis of the coalescence-induced jumping including the derivation of the present model and comparisons with the existing models (Section S5); and visualization of spatially-confined growth, coalescence-induced jumping, jumping relay and sequential multi-droplet coalescence

(Section S6 and Videos S1-S6).

Video S1: In the nucleation period, most nuclei regularly distributed within the surface roughness rather than atop the structure, showing a spatial preference of nucleation sites

Video S2: The spatially confined growth of a squeezed droplet leads to various states, including the impaled Wenzel state with a trapped droplet body, the suspended Cassis state/dewetting transition and a spontaneous ejection from the surface.

Video S3: Different modes of the Wenzel-Wenzel coalescence events, including the Wenzel state, the Cassie state (dewetting transition) and the jumping.

Video S4: Different modes of the Wenzel-Cassie coalescence events., including the Wenzel state and the dewetting transition.

Video S4: A flying droplet generated from the coalescence or the spatially confined growth touch and assimilate other sessile Wenzel or Cassie droplets around its flight path, and leads to a jumping relay or a dewetting transition.

Video S5: The nanoscale Cassie-Cassie coalescence events lead to the spontaneous jumping or a larger Cassie droplet.

Video S6: A 9-droplet coalescence event, the sequential coalescence not only changes the droplet size but also triggers the wetting state transition.

**Declaration of interests**

The authors declare no competing financial interests.